\documentclass[
journal=ancac3, 
manuscript=article, layout=twocolumn]{achemso}
\usepackage[version=3]{mhchem} 
\usepackage{siunitx}
\usepackage{xcolor}
\usepackage[font=footnotesize,labelfont=bf]{caption}
\usepackage{booktabs}

\newcommand{\RM}[1]{\MakeUppercase{\romannumeral #1{}}}

\usepackage{setspace}
\author{Laura Katharina Scarbath-Evers }
\affiliation{Martin-Luther University Halle-Wittenberg, Institute of Chemistry, Halle/Saale, Germany
}
\alsoaffiliation{\textit{Current affiliation: Schr\"odinger Gmbh, Q7 23, 68161 Mannheim, Germany}}
\altaffiliation{Contributed equally to this work}

\author{Ren\'e Hammer }
\affiliation{Martin-Luther University Halle-Wittenberg, Institute of Physics, Halle/Saale, Germany
}
\altaffiliation{Contributed equally to this work}

\author{Dorothea Golze}
\affiliation{Department of Applied Physics, Aalto University School of Science, FI-00076 Aalto, Finland
}
\author{Martin Brehm}
\affiliation{Martin-Luther University Halle-Wittenberg, Institute of Chemistry, Halle/Saale, Germany
}

\author{Daniel Sebastiani}
\affiliation{Martin-Luther University Halle-Wittenberg, Institute of Chemistry, Halle/Saale, Germany
}
\author{Wolf Widdra}
\affiliation{Martin-Luther University Halle-Wittenberg, Institute of Physics, Halle/Saale, Germany
}
\email{wolf.widdra@physik.uni-halle.de}

\let\oldmaketitle\maketitle
\let\maketitle\relax

\title[\texttt{achemso} demonstration]
{From Flat to Tilted: Gradual Interfaces in Organic Thin Film Growth}
\begin{document}
\twocolumn[
\begin{@twocolumnfalse}
\oldmaketitle

\begin{abstract}
We investigate domain formation and local morphology of thin films of $\alpha$-sexithiophene ($\alpha$-6T) on Au(100) beyond monolayer coverage by combining high resolution scanning tunneling microscopy (STM) experiments with electronic structure theory calculations and computational structure search. We report a layerwise growth of highly-ordered enantiopure domains. 
For the second and third layer, we show that the molecular orbitals of individual $\alpha$-6T molecules can be well resolved by STM, providing access to detailed information on the molecular orientation.  We find that already in the second layer the molecules abandon the flat adsorption structure of the monolayer and  adopt a tilted conformation. Although the observed tilted arrangement resembles the orientation of $\alpha$-6T in the bulk, the observed morphology does not yet correspond to a well-defined surface of the $\alpha$-6T bulk structure. A similar behavior is found for the third layer indicating a growth mechanism where the bulk structure is gradually adopted over several layers. \\
\end{abstract}
\end{@twocolumnfalse}

]


\renewcommand*\rmdefault{bch}\normalfont\upshape
\rmfamily
\section*{}
\vspace{-1cm}





\section{Introduction}
During the last decades, small molecular weight organic semiconducting materials have received significant attention,  due to their low production costs, biocompatibility,\cite{Rivnay2014} structural variety,\cite{Anthony2006,Murphy2007,Yamada2008} and tunability for rational design approaches.\cite{Mei2013,Dey2019}
$\alpha$-\textit{trans}-sexithiophene, further denoted as $\alpha$-6T, is a prominent representative of the oligothiophene family. Oligothiophenes are organic semiconductors with good hole conducting properties \cite{Dodabalapur1995,Garnier1993a,Kan2015} and real life application in organic field effect transistors (OFETs).\cite{Ong2004,Horowitz1989,Horowitz1990,Mannebach2013} Morphology and electronic properties of $\alpha$-6T on metal surfaces have been investigated from single molecule adsorption \cite{Scarbath-Evers2019} to structural assembly and domain formation of monolayers.\cite{Kiguchi2004a,Kiguchi2004,Yoshikawa2004, Maekinen2005,Kiel2007,Duncker2008,Hoefer2011}\par
\begin{figure*}[htbp]
\includegraphics[width=0.99\linewidth]{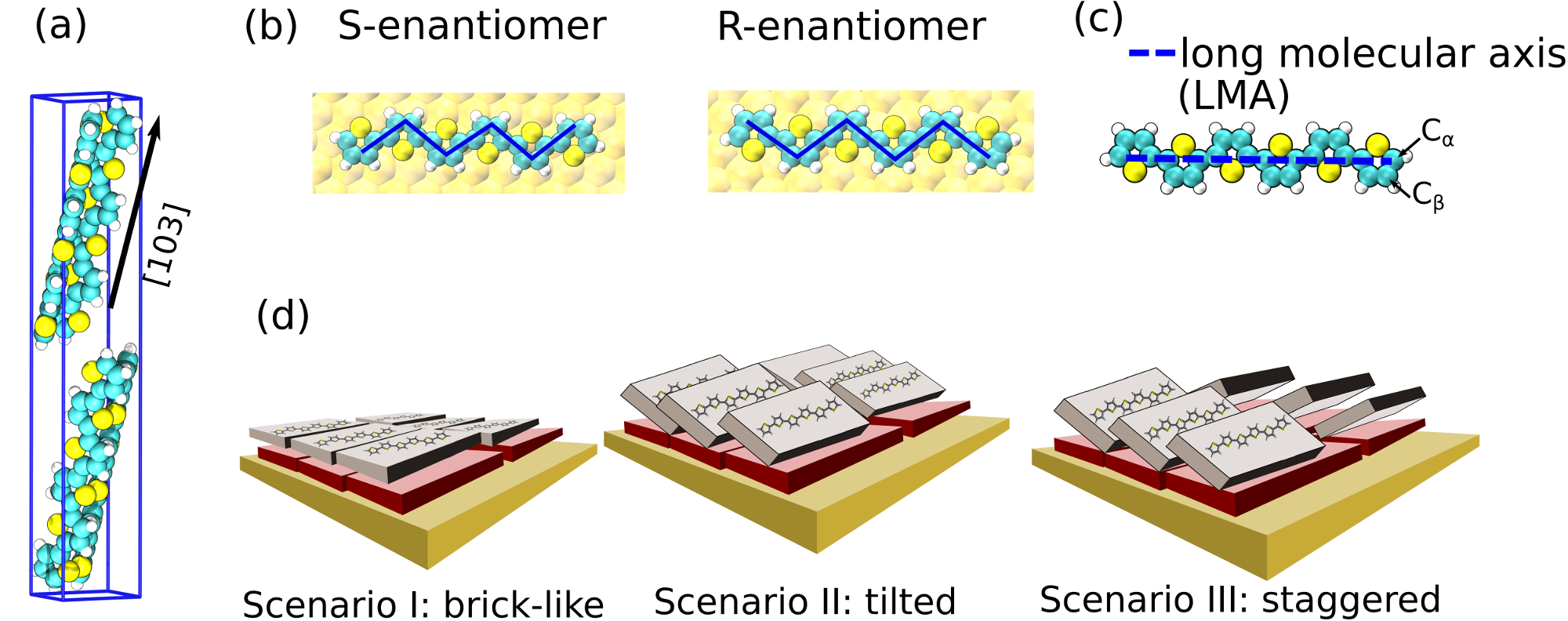}
\caption{(a) Low temperature crystal structure of $\alpha$-6T.\cite{Horowitz1995} (b) Enantiomeric forms of $\alpha$-6T after flat adsorption on a planar surface. (c) Visualization of the long molecular axis and the C$_{\alpha}$ and C$_{\beta}$ carbon atoms. (d) Schematic of possible scenarios for the growth of $\alpha$-6T beyond the monolayer: flat-lying molecules (brick-like, scenario I), equally tilted molecules (tilted, scenario \RM{2}) or a staggered arrangement (staggered, scenario \RM{3}). }
\label{comic_scenario}
\end{figure*}

The formation of flat monolayers has been found on all metal surfaces. Due to the prochirality of $\alpha$-6T in its \textit{all-trans} conformation, such a flat adsorption leads to the formation of two enantiomeric forms, the S and the R enantiomer (see Figure \ref{comic_scenario} (b)). Note that the enantiomeric structure comprises the entire molecule-surface system, and not only the molecule. For monolayer coverage, formation of extended chiral domains that consist exclusively either of the S enantiomer or the R enantiomer has been reported for $\alpha$-6T on Ag(100), \cite{Duncker2008} Ag(110), \cite{Wagner2011} Au(100), \cite{Hoefer2011} and Au(111). \cite{Kiel2007}
A commonly discussed driving force for the formation of homochiral domains is the minimization of steric repulsion and the subsequent increase in packing density; for a detailed discussion see Ref.~\citenum{Kiel2007}.
However, heterochiral monolayers consisting of R and S enantiomers have also been observed after thermally induced cis-trans-isomerization of $\alpha$-6T molecules on Ag(100). \cite{Duncker2008}\par

$\alpha$-6T crystallizes in its low-temperature structure\cite{Horowitz1995,Hermet2005} depicted in Figure~\ref{comic_scenario} (a) upon sublimation at low pressure. Since our experimental setup matches these conditions, our discussion refers always to the low-temperature phase. In the bulk, the molecules are packed in a staggered herringbone fashion characteristic for many rod-shaped molecules.\cite{Baker1993,baker_knachel_fratini_adams_1988,Athouel1996,Resel2001}  Investigations of thick films of $\alpha$-6T on organic and inorganic substrates found that $\alpha$-6T eventually adopts its herringbone bulk structure.\cite{Simbrunner2013,Koller2007}
However, the transition from the flat lying monolayer found on metallic surfaces to the bulk structure is not yet understood.\par
The determination of the structure of organic thin films might be further impeded by the potential occurrence of surface induced polymorphism (SIP). It becomes likely if several structural arrangements with similar free energies are coexisting and a structural optimization is impeded by kinetic barriers.\cite{Jones2016} Indeed for $\alpha$-6T, SIP has been reported after growth with high deposition rates.\cite{Servet1993,Servet1994,Moser2013} However, the low-deposition rates used here lead to well-defined phases instead.
 \\
The molecular arrangement and packing of $\alpha$-6T in the layers close to the surface determine the opto-electronic properties and device performance. After injection at the metallic interface, the electron-hole
transport through the organic layers depends strongly on the local chemical interactions.\cite{Bromley2004,Bredas2004,Heimel2008} A thorough understanding of the morphology of thin films is therefore of paramount importance for optimization of organic devices.\par 
In the following, we consider three different growth scenarios for $\alpha$-6T on top of the first flat monolayer formed on Au(100).
I: The enforced, flat adsorption structure of the first layer is still retained in the second layer as schematically shown in Figure~\ref{comic_scenario} (d). A layer-by-layer growth of thin  ordered two-dimensional films is a common motif for small organic semiconductors \cite{Wu2016,Liu2016,Zhang2016} and has  been, e.g., observed for perylenetetracarboxylic dianhydride (PTCDA) on Ag(111) at elevated temperatures \cite{Kilian2004} and suggested for $\alpha$-6T on Ag(100) based on molecular dynamics simulations.\cite{Chen2009} A flat second layer has also been found for $\alpha$-6T on Ag(110) by means of scanning tunneling
microscopy (STM) measurements \cite{Wagner2011} as well as on Au(111) based on photoelectron emission microscopy (PEEM). \cite{Bronsch2018}\par
II: The flat structure is abandoned and the growth of $\alpha$-6T continues in its low temperature bulk structure with the (010) oriented surface as contact plane. The formation of $\alpha$-6T crystals with (010) orientation on top of a flat monolayer has been observed for $\alpha$-6T on TiO$_2$(110).\cite{Haber2008} Generally, the bulk (010) surface has been reported as contact plane for the adsorption of $\alpha$-6T on substrates, whose interactions with the molecules are weaker than for metallic surfaces but still sufficiently strong to favour a quasi-horizontal alignment of the molecules.\cite{Ivanco2007,Koini2009} Adsorption in (010) orientation implies that all molecules directly adsorbed on top of the monolayer are tilted by the same angle $\phi$ with respect to the surface plane, as shown in Figure~\ref{comic_scenario} (d). For this scenario we expect the formation of 3D islands, rather than a layer-by-layer growth.\cite{Wagner2011,Bronsch2018,Haber2008} Alternatively, formation of  $\alpha$-6T crystals with a
(001) contact plane on top of the flat monolayer would lead to a quasi-standing molecular orientation. Such a vertical arrangement has been observed for organic semiconductors on weakly interacting substrates.\cite{Opitz2016} However, here our STM results disregard this structure easily.\par
Instead of the expected growth scenarios, \RM{1} or \RM{2}, we find that the molecules in the second layer are aligned in a staggered arrangement (scenario \RM{3} in Figure~\ref{comic_scenario} (d)), i.e., the molecules of adjacent rows are tilted in the opposite directions.
These findings indicate a growth mechanism in which the molecular arrangement gradually approaches the bulk structure over several layers, which is further corroborated by orbital-resolved STM images of a fully formed third layer.

\par
 We characterize the structure of the $\alpha$-6T bilayer and molecules adsorbed on top of it  by orbital-resolved STM images and computational data. Our computational model is an image-charge augmented hybrid quantum mechanics/molecular mechanics (QM/MM) scheme, which has been recently developed for the simulation of adsorption processes at metallic interfaces.\cite{Golze2013} The $\alpha$-6T layers are treated by density functional theory (DFT), while the metal and the interactions between metal and molecules are described at the MM level of theory, see Experimental and Computational Details.\par

\section{Results and Discussion}

\subsection{Characterization of the bilayer domains}
\begin{figure*}[htbp]
\includegraphics[width=0.99\linewidth]{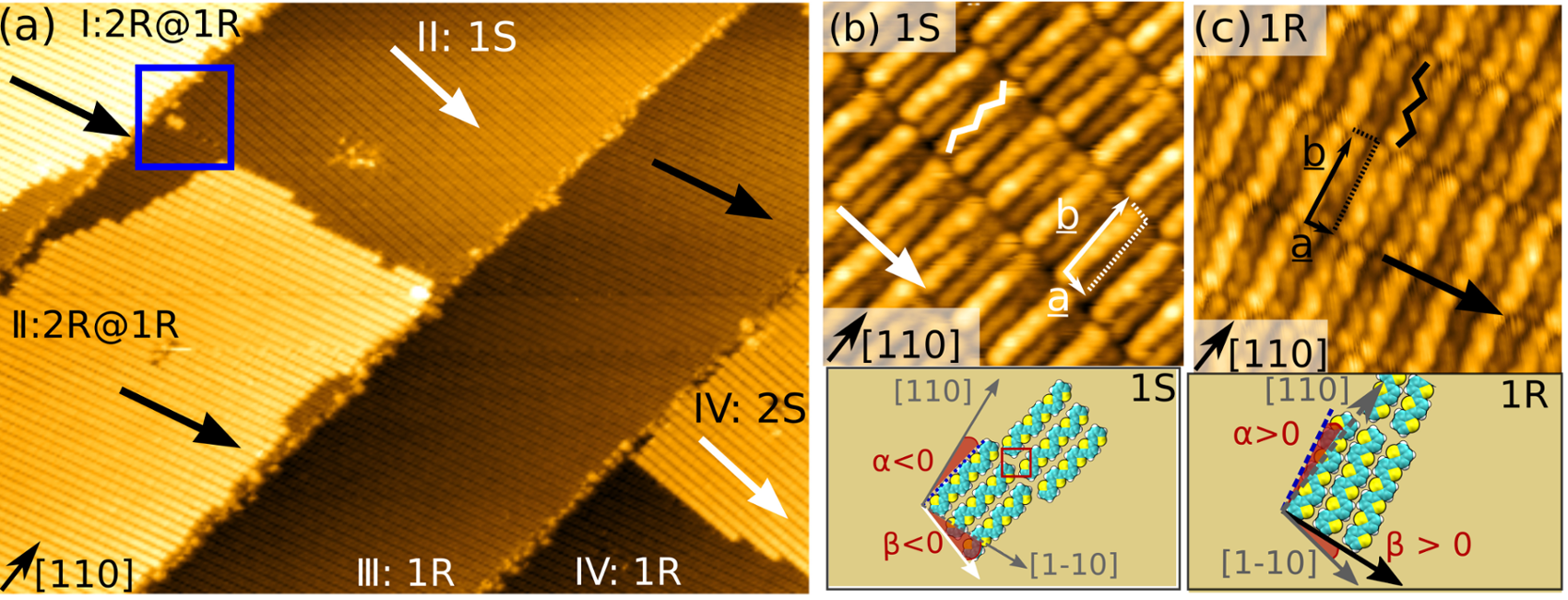}
 \caption{(a) 1.5 monolayer-thick film of $\alpha$-6T (I$_{\rm{T}}=\SI{10}{\pico\ampere}$, U$_{\rm T}=\SI{2.2}{\V},  \; \SI{80}{\kelvin}$) on four Au(100) terraces (\RM{1}-\RM{4}). The growth of enantiopure R-domains (2R) and S-domains (2S) on top of the first layer can be observed on terrace \RM{1}, \RM{2} and, \RM{4} respectively. (b) High-resolution STM image of a monolayer domain with S and (c) R chirality (I$_{\rm{T}} = \SI{20}{\pico\ampere}$, U$_{\rm{T}} =
\SI{2.3}{\V}, \SI{80}{\kelvin}$). The angles $\alpha$ and $\beta$ in the schematic characterize the growth direction of the domains.  }
\label{bl_exp}
\end{figure*}
In the following, we analyze the growth of extended domains of $\alpha$-6T on Au(100) beyond the monolayer coverage in order to distinguish between a layer-by-layer growth (scenario \RM{1}) and a Stranski-Krastanov growth (scenario \RM{2}). Figure \ref{bl_exp} (a) shows a large-scale STM image of a 1.5 monolayer-thick film of $\alpha$-6T on Au(100) on four different substrate terraces (\RM{1}-\RM{4}). 
On terrace \RM{2} and \RM{3}, the growth of densely packed monolayer domains of $\alpha$-6T can be observed which are further denoted as 1S (terrace \RM{2}) and 1R (terrace \RM{3}). For both monolayer domains, the number of molecules per area amounts to 0.58 molecules per $\si{nm^2}$.
 
The growth directions of the molecular rows is different for each of the two monolayer domains and can be characterized by the angles $\alpha$ and $\beta$ depicted in the insets in Figure~\ref{bl_exp} (b) and (c). The angle $\alpha$ encloses the [110] direction and the long molecular axis (LMA) of the molecule (blue dotted line), whereas $\beta$ is the angle between the growth direction of the molecular rows (indicated by a black/white arrow) and the [$1\bar{1}0$] direction. In the 1S domain, the angles $\alpha$ and $\beta$ are  $<0^{\circ}$, while they are are both $>0^{\circ}$ in the 1R domain. The absolute values of $\alpha$ and $\beta$ are between $5^{\circ}-10^\circ$, i.e., the LMA is in both domains almost parallel to the [110] direction and almost perpendicular to the direction of the molecular rows. These findings are consistent with those of an extensive study of the $\alpha$-6T monolayer on Au(100) as discussed in Ref.~\citenum{Hoefer2011}. \par
Figures~\ref{bl_exp} (b) and (c) show a high-resolution image of an 1S and 1R domain, respectively (see Figure~S1 in the supporting information (SI) for a large-scale image). The molecules are clearly resolved and appear alternately brighter and darker. This variation in contrast is due to the height corrugation of the underlying reconstructed gold surface, see also Ref. \citenum{Hoefer2011}. A fine structure of six bright protrusions within each molecule is visible, where each protrusion corresponds to one of the six thiophene units of the molecule. The protrusions are arranged in a zig-zag pattern exhibiting the same orientation within one domain. Comparing 1S and 1R domain, the zig-zag patterns are oriented in the opposite direction. In the 1S domain the pattern can be assigned to an adsorbed $\alpha$-6T molecule in the S-enantiomeric and in the 1R domain in the  R-enatiomeric form.\cite{Duncker2008} Each domain accommodates exclusively one of the enantiomeric forms.  \par
In both domains the surface unit cell contains exactly one molecule. Note that we do not account here  for the substrate buckling discussed above. The absolute values of the unit cell vectors are very similar, deviating by only 0.3~{\AA}, see Table~\ref{tab1} for the cell parameters. The different orientation of the molecular rows in the two domains, however, is reflected in different cell angles of the surface unit cell. In the 1S domain, the unit cell angle is smaller than $\SI{90}{\degree}$ (roughly $\SI{87}{\degree}$), whereas in the 1R domain the unit cell angle deviates with $\SI{95}{\degree}$ also slightly from a rectangular geometry. The deviation of the cell angle from $\SI{90}{\degree}$ leads to a small translation of the molecules in the adjacent row. As a result, the terminal C$_{\alpha}$ atoms of the molecule in one row are placed between two adjacent molecules in the next row, which is highlighted by a red rectangle in the inset of Figure~\ref{bl_exp} (b). This small translation of the molecular rows within one layer, further denoted as $\Delta_{\textrm{row}}^{\textrm{intra}}$, facilitates a closer contact between molecules of two adjacent rows and hence increases the packing density, a concept discussed for $\alpha$-6T on Au(111) earlier.\cite{Kiel2007}\par
The growth of a second layer is observed on terraces \RM{1}, \RM{2} and \RM{4}. 
With an apparent height of $\approx \SI{3.1}{\angstrom}$, the domains that grow on the monolayer consist only of one layer of molecules.  No  three-dimensional islands can be observed.
 This is a clear indication for a layer-by-layer growth.
Figure~\ref{bl_exp}~(a) yields also insight into the growth behavior at step edges and domain boundaries. The growth of the bilayer on terrace \RM{2} stops at step edges to terrace \RM{1} and \RM{3}. A similar behavior is observed for the bilayer on terrace \RM{4}. The blue rectangle in Figure~\ref{bl_exp} (a) encloses the boundary between the 1R and 1S domains on terrace \RM{2}. It is visible that the growth of the bilayer is restricted to the 1R domain and does not extend beyond the domain border.\par

\begin{table}[htbp]
\caption{Measured lattice constants [$\si{\angstrom}$] and angle [$\si{\degree}$] of the unit cell of monolayer and bilayers of $\alpha$-6T on Au(100) in comparison with lattice constants of the $\alpha$-6T-(010) surface.}
\label{tab1}
\centering
\begin{tabular}{cccc}
\hline
Structure & $|b|$ &  $|a|$ & $\gamma$ \\
\hline
1S    & 25.9 & 6.5 & $87 \pm 1$ \\
1R    & 25.9 & 6.8 &  $95 \pm 1$ \\
2S@1S & 51.8 &6.5 &  $87 \pm 1$ \\
2R@1R & 51.8 & 6.8 &  $95 \pm 1$ \\
6T(010) \cite{Horowitz1995,Oehzelt2009} & 44.708 & 6.029 &  89.4 \\

\hline
\end{tabular}
\end{table}

\begin{figure*}[htbp]
\includegraphics[width=0.99\linewidth]{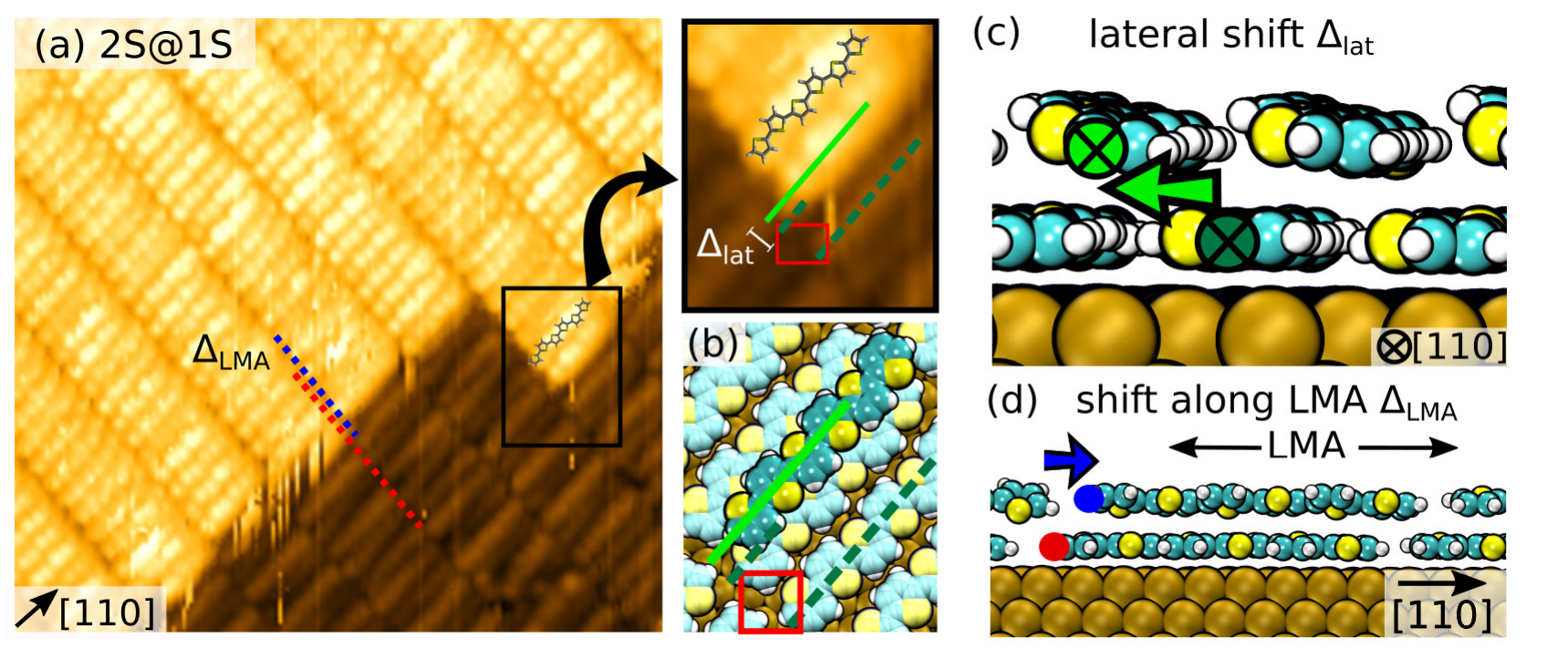}
\caption{ (a) STM image (I$_{\rm{T}}=\SI{10}{\pico\ampere}$, U$_{\rm T}=\SI{2.2}{\V},  \; \SI{80}{\kelvin}$) of a 2S domain (bright) on top of a 1S domain (dark). The lateral shift, $\Delta_{\textrm{lat}}$, of the 2S molecules with respect to the first layer is indicated by dark and light green lines. The shift along the LMA, $\Delta_{\textrm{LMA}}$, is indicated by the red and blue lines. (b)-(d) Computed 2S@1S structures. (b) Top view restricting the visualization to one 2S molecule to display $\Delta_{\textrm{lat}}$. (c) View perpendicular to the LMA indicating  $\Delta_{\textrm{lat}}$ by green dots. (d) View perpendicular to the LMA indicating $\Delta_{\textrm{LMA}}$ by blue and red dots. }
\label{bl_shift}
\end{figure*}

Similar to the first layer, the molecules in the second layer assemble in two enantiomeric pure domains that can be distinguished by the angles $\alpha$ and $\beta$ characterizing the orientation of the individual molecules and that of the molecular rows. Both angles are similar to those of the first layer. For the second layer on terrace \RM{1} and \RM{2}, $\alpha$ and $\beta$ are similar to that of the underlying 1R domain. This implies that the second layer consists of $\alpha$-6T molecules in the R-enantiomeric form, which we denote by 2R. In the following, we adopt the notation 2R$@$1R for a 2R layer growing on top of a 1R layer, where we want to emphasize the relation between the layers. On terrace \RM{4}, the growth direction of the second layer is the same as for the 1S domain on terrace \RM{2}. Evidently, the second layer is of S chirality (2S). For this terrace, it is not possible to determine the chirality of  the underlying monolayer domain since the domain boundaries in the first layer are not visible. However, we observed only enantiomeric pure bilayers, i.e., 2R@1R and 2S@1S structures, see also Figure~\ref{bl_shift} (a). There is no evidence of $\alpha$-6T bilayers, where the first and second layer have a different chirality. The parameters of the unit cell for the bilayer structures are reported in Table~\ref{tab1}. The latter are identical to the parameters for the monolayer, except that the absolute value of cell vector $\mathbf{b}$ is increased by a factor of 2. We will return to the two-fold increase of the cell vector when analyzing the molecular orientation.

\subsection{Position of second-layer molecules} 
In the following, we discuss the relative position of the molecules in the second layer with respect to the first layer by the example of a bilayer structure with S chirality. Figure~\ref{bl_shift} (a) depicts a high-resolution STM image of a 1.5 monolayer-thick film of a 2S domain growing on top of a 1S layer. The corresponding computational 2S@1S structure is shown in Figure~\ref{bl_shift} (b-d).\par

\begin{table*}[h!]
\caption{Structural and energetic characterization of the bilayer structures comparing computation (comp) and experiment (exp). $\Delta E \; (\si{\kilo\joule\per\mol})$ is the energy difference per molecule with respect to the 2S$@$1S structure with staggered configuration, i.e., $\Delta E$ = $(E_{\textrm{structure}}$ - $E_{\textrm{2S@1S(\RM{3})}})/n$ where $n$ is the number of molecules in the simulation cell.  $\Delta_{\textrm{LMA}}$ ($\si{\angstrom}$) is the median shift along the LMA, $\Delta_{\textrm{lat}}$ ($\si{\angstrom}$) the median shift in lateral direction, and $\phi$ ($\si{\degree}$) is the median value of the tilt angle between molecular and surface plane of the molecules in the second layer. $\Delta_{\textrm{row}}^{\textrm{intra}}$ ($\si{\angstrom}$) is the translation of molecular rows within the first layer. The values for the $\alpha$-6T bulk were derived from the experimentally measured $\alpha$-6T low temperature structure.\cite{Horowitz1995} 
}
\label{bl_features}
\centering
\begin{tabular}{cccccccccc}
\toprule
  &  & $\Delta E$  & \multicolumn{2}{c}{$\Delta_{\textrm{lat}}$}  & \multicolumn{2}{c}{ $\Delta_{\textrm{LMA}}$}  &  \multicolumn{2}{c}{ $\phi$} &  $\Delta_{\textrm{row}}^{\textrm{intra}} $  \\\cmidrule(l{0.4em}r{0.4em}){3-3}\cmidrule(l{0.4em}r{0.8em}){4-5}  \cmidrule(l{0.4em}r{0.8em}){6-7} \cmidrule(l{0.4em}r{0.8em}){8-9}\cmidrule(l{0.4em}r{0.4em}){10-10}
  structure & scenario & comp & comp & exp  & comp & exp & comp & exp & comp\\\midrule


 2S$@$1S  & \RM{3} &  0.0 & 2.5 & 3  & 2.3   & 3 & 14 & - & 0.8 \\
 2R$@$1R  & \RM{3} &  -0.5   & 2.6 &  3   & 2.4  & 3  & 12 & - &0.7 \\
 2S$@$1S  & \RM{2} &  -0.1   & 2.2 &  -    & 0.2  & -  & 13 & - & 2.3 \\
 $\alpha$-6T bulk & \RM{2}-like  & - & 2.7 & 2.7 & 1.1 & 1.2 & 30 & 33 \cite{Oehzelt2009}  & -   \\
\bottomrule
\end{tabular}
\end{table*}
As discussed in the previous section, molecules in two consecutive rows in the monolayer are not perfectly aligned but slightly translated by a shift $\Delta_{\textrm{row}}^{\textrm{intra}}$, which is indicated by a red rectangle in the inset in Figure~\ref{bl_shift}~(a) and  Figure~\ref{bl_shift}~(b). The translation $\Delta_{\textrm{row}}^{\textrm{intra}}$ 
occurs also in the second layer. If the 2S molecules were located directly on top of the 1S molecules, we would also expect a $\Delta_{\textrm{row}}^{\textrm{intra}}$ shift at the edge between 1S and 2S layer parallel to the LMA. The expected position of the first molecular row in the 2S layer is visualized by dark green, dashed lines in Figure \ref{bl_shift} (a) and (b). Comparing the positions of the molecules in the first and in the second layer of two adjacent rows, however, we find that they are directly aligned lacking the expected translation in row direction. The actual position is indicated by a bright green line in the inset in Figure~\ref{bl_shift}~(a) and Figure~\ref{bl_shift}~(b). This implies that the molecules in the second layer are not directly on top of the molecules in the first layer, but laterally shifted. We denote this shift in the following as $\Delta_{\textrm{lat}}$.
 At the 1S/2S edge, we measured a shift in the periodicity of $|a|/2$, which corresponds to $\Delta_{\textrm{lat}} \approx \SI{3}{\angstrom}$. Note that $\Delta_{\textrm{lat}}$ can only be roughly determined from the experimental STM image since the tilted arrangement of the 2S molecules (see next section) should lead to a slight off-set of the long axis of the bright rods shown in Figure~\ref{bl_shift}~(a) with respect to the actual LMA. We find a lateral shift between 1S and 2S layer also in our calculations, directly visible from Figure \ref{bl_shift}~(c), where a view along the molecular rows is displayed for the computed 2S$@$1S structure. The calculated median value of $\Delta_{\textrm{lat}}$ agrees with 2.5~{\AA} well with the experimental estimate.\par

Aside from the the lateral shift, $\Delta_{\textrm{lat}}$, the molecules in the second layer are also translated in direction of the LMA as indicated by the distance between the red and blue dotted lines in Figure \ref{bl_shift} (a). The red dotted line marks the separation between two molecular rows in the first layer and the blue dotted line marks the separation in the second layer. The shift along the LMA, $\Delta_{\textrm{LMA}}$, obtained from experiment is approximately $\SI{3}{\angstrom}$. The $\Delta_{\textrm{LMA}}$ shift in the computed structure is shown in Figure \ref{bl_shift} (d) depicting the molecules with their LMA parallel to the paper plane. The separation between the 1S and 2S rows is depicted with red and blue circles, respectively. The median value of $\Delta_{\textrm{LMA}}$ is $\SI{2.3}{\angstrom}$, well in agreement with the experimental value.\par
The shifts $\Delta_{\textrm{LMA}}$ and $\Delta_{\textrm{lat}}$ are also present in the bilayer of opposite chirality, 2R$@$1R (see Table \ref{bl_features}) which indicates that those two shifts are universal, characteristic features of the $\alpha$-6T second layer regardless of the chirality of the domains.

\subsection{Molecular orientation in the second layer}
\label{subsection:tilt}

High-resolution STM measurements in combination with computational results point strongly towards a staggered arrangement of $\alpha$-6T in the second layer (scenario \RM{3} in Figure~\ref{comic_scenario} (d)), which is now discussed in detail.\par
\begin{figure*}[t!]
\centering
\includegraphics[width=0.9\linewidth]{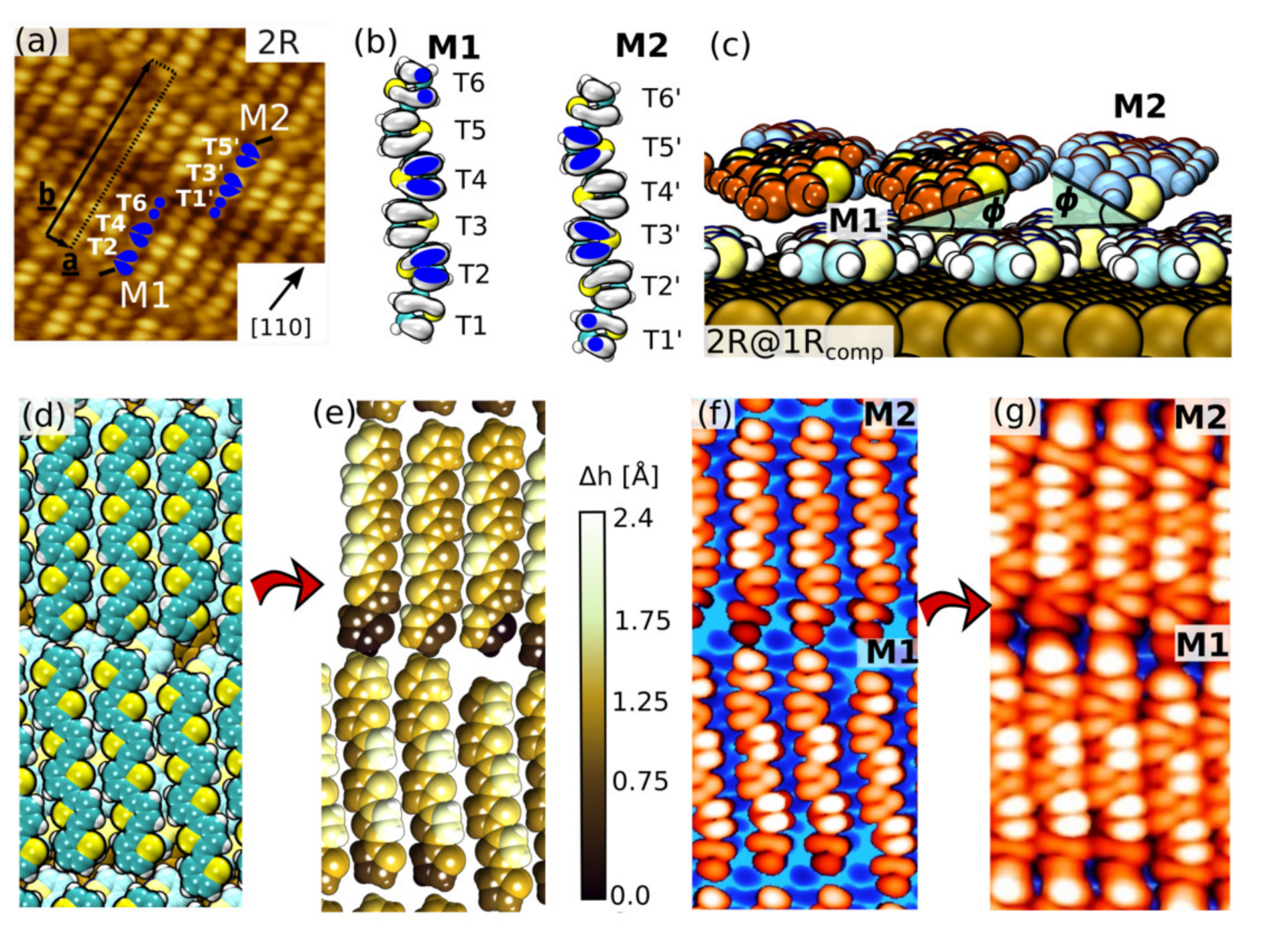}
\caption{(a) High-resolution STM image of a second layer with R chirality  measured at a bias voltage of $-$1.35~V (I$_{\rm{T}}=\SI{10}{\pico\ampere}$,  \; $\SI{80}{\kelvin}$) showing a protrusion pattern and two molecules in the unit cell labeled as M1 and M2. (b) Computed density of the HOMO for an isolated $\alpha$-6T molecule. The molecule is oriented as in the 2R structure with the sulfur of the first thiophene unit (T1, T1') pointing to the right. (c) Computationally optimized 2R$@$1R structure exhibiting a staggered arrangement as in scenario \RM{3}. (d) Top-view of the optimized bilayer and (e) corresponding color-coded profile of the relative height of the atoms in the second layer. The atom closest to the monolayer is set as reference. (f,g) Calculated STM image of the 2R$@$1R structure plotted at a larger (d) and smaller (e) isovalue. }
\label{bl_tilt}
\end{figure*}

The high resolution STM image of a 2R island displayed in Figure~\ref{bl_tilt}~(a) shows a distinct protrusion pattern for the molecules in the second layer. Details of the molecular electronic structure are visible since the first layer decouples the electronic structure of the second layer from the surface.  This is a known effect exploited previously by using NaCl films for imaging molecular orbitals.\cite{Repp2005,Cavar2005} On the contrary, high resolution images of molecules in the monolayer show featureless, rod-like molecules since the direct contact with the metal surface leads to a perturbation and broadening of the molecular electronic states. One molecule in the second layer is represented by six droplet-shaped, well separated protrusions of different brightness followed by a blurry extension. In Figure \ref{bl_tilt} (a), the protrusions are indicated by blue marks and the blurry extensions by a black line. The six droplet-shaped protrusions are arranged in three pairs. The brightness pattern of the protrusions and the blurry extension extend the periodicity of the STM from one to two molecules per unit cell which are denoted in the following as M1 and M2. This is the first indication that molecules are staggered as in scenario \RM{3}. For structures \RM{1} and \RM{2}, the unit cell should contain only one molecule. \par
In order to assign the protrusion pattern in the experimental STM image to structural motifs in the molecules,
we calculate the probability density  ($ \| \Phi_{\textrm{HOMO}} \| ^2$) of the highest occupied molecular orbital (HOMO). The distribution of the electron density of the HOMO of an isolated $\alpha$-6T molecule is displayed in Figure~\ref{bl_tilt} (b) and exhibits several nodal planes: one within the molecular plane and planes perpendicular to the latter passing through the center of the thiophene units and through the interring bridges. In the following, we denote the six thiophene units for molecules M1 with T1-T6 and for molecule M2 with T1'-T6', see Figure~\ref{bl_tilt} (b). For each thiophene unit we observe a pair of droplet-shaped lobes.
The electron density is mainly located at the aromatic carbon atoms and significantly reduced at the sulfur atoms. This observation is discussed in terms of the electronic structure of a single thiophene molecule as well as the oligomers in more detail in the SI. 
For molecule M1, the density distribution of units T2, T4 and T6 resembles in symmetry and shape the protrusion pattern in the experimental STM, while for molecule M2 the resemblance is found for units T1', T3' and T5'. The other thiophene units are not resolved in the experimental image. \par

The alternating pattern of visible and non-visible thiophene units in the STM image can be explained by height differences between the C$_{\beta}$ atoms of adjacent units, which we will rationalize in the following based on our theoretical results (see Figure~\ref{comic_scenario} (c) for nomenclature). 
Our computational structure search found indeed a tilted, staggered arrangement as shown in Figure~\ref{bl_tilt}~(c). The median computed tilt angle $\phi$ is $\SI{12}{\degree}$ and the height profile of the molecules in the second layer is presented in Figure~\ref{bl_tilt}~(e). For molecule M1, the C$_{\beta}$ atoms of units T2, T4 and T6 point upwards and the sulfur atoms downwards, whereas the opposite holds for units T1, T3 and T5. Molecule M2 is tilted in the opposite direction and C$_{\beta}$ atoms of T1', T3' and T5' are directed away from the surface. The height difference $\Delta$h between the C$_{\beta}$ atoms of neighboring units is on average 0.7~{\AA}. Since the HOMO has the largest density at the C$_{\beta}$ atoms and due to the strong decrease in tunneling probability with increasing distance, the HOMO is only visible for the units where the C$_{\beta}$ atoms point towards the STM tip. This is directly confirmed by the computed STM image shown in Figure~\ref{bl_tilt} (f). The units with C$_{\beta}$ atoms tilted upwards appear as a bright pair of droplet-shaped lobes. The other units are significantly darker.\par
 The protrusions in the experimental image that we assigned to the units T6 and T1' appear on average darker than the other protrusions. We observe this variation in brightness to some extent also in our computational results. Some of the T6 and T1' units are bent towards the surface as evident from the height profile in Figure~\ref{bl_tilt} (e). As a result, they are noticeably darker in the STM. This becomes more obvious when plotting the computed STM at lower current to facilitate comparison to experiment, see Figure~\ref{bl_tilt} (g). As in the experimental image, the units T2, T4, T3' and T5' dominate in brightness. Furthermore, they superimpose the shape of the lobes originating from the thiophene units where the C$_{\beta}$ atoms point downwards. In addition, the lobes of the terminal units T1 and T6' have no longer a distinct shape resembling the blurry extension in Figure~\ref{bl_tilt} (a).\par
 In summary, the following four points are strong evidence that the molecules in the second layer are arranged in a staggered configuration (scenario \RM{3}): i) We have two molecules in the unit cell. ii) The computed HOMO density of the $\alpha$-6T molecule shows droplet-shaped lobes organized in pairs, which is also observed in experiment. For a flat structure six pairs are expected. However, only three out of six appear in the STM. iii) The structure optimization yields a staggered arrangement and iv) the corresponding computed STM image resembles strongly the experimental image, in particular, when plotted at lower currents. Note that the electronic structure of the metal is not explicitly accounted for in our computational model, which confirms that the second layer is largely decoupled from the metallic surface.\par
Our computational optimization procedure yields also a staggered structure for the 2S$@$1S bilayer. The structural features are similar to 2R$@$1R, see Table~\ref{bl_features}. Interestingly, we also found a 2S$@$1S structure with a configuration as in scenario \RM{2}, i.e., the molecules in the second layer are always tilted in the same direction, see Figure~S3 (SI). Both structures, 2S$@$1S-\RM{2} and 2S$@$1S-\RM{3}, are equal in energy. For 2R$@$1R, a structure of type \RM{2} has not been obtained. Furthermore, there is no experimental evidence for equally tilted arrangements for any of the bilayers. It might be indeed that structures \RM{2} and \RM{3} are similar in energy. In fact, scenario \RM{2} resembles more closely the $\alpha$-6T bulk structure than \RM{3}, as discussed in the next section. However, our computational model applies periodic boundary conditions and is too small to capture domain effects. Moreover, the electronic structure of the metal is not taken explicitly into account in our QM/MM approach neglecting possible charge transfer between bilayer and metal. Both approximations might tip the energy balance towards structure \RM{3}. \par

\begin{figure*}[htbp]
\includegraphics[width=0.95\linewidth]{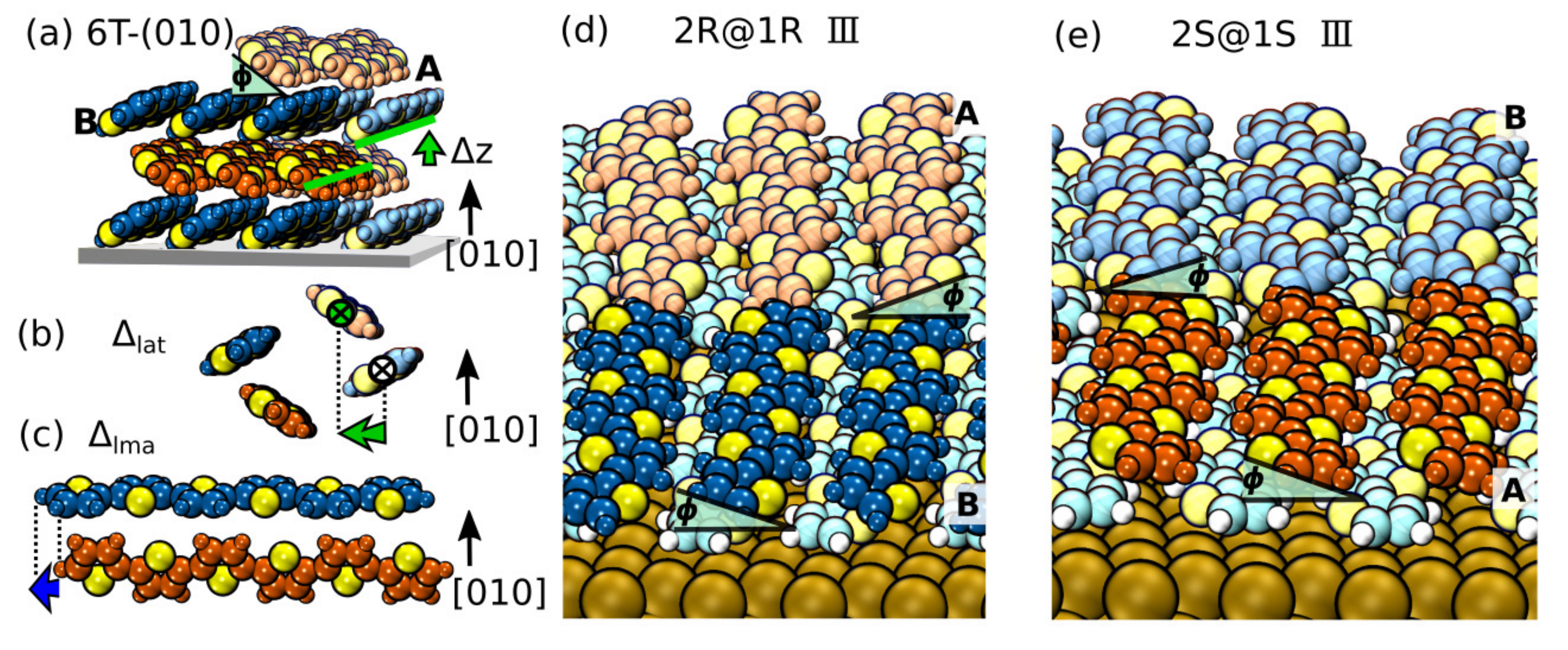}
\caption{  (a) Experimental crystal structure of the $\alpha$-6T bulk \cite{Horowitz1995} in the (010) terminated orientation. The (010) plane is sketched in grey. Indicated are the tilt angle $\phi$ and the vertical translation, $\Delta z$ between two molecules with opposite tilt (green lines). Visualization of the (b) lateral shift ($\Delta_{\textrm{lat}}$) and (c) shift along the LMA ($\Delta_{\textrm{LMA}}$) between layers A and B in the bulk structure. Staggered bilayer structures (d) 2R$@$1R and (e) 2S$@$1S obtained from the computational structure search. }
\label{b_vs_c}
\end{figure*}


\subsection{Comparison of the second layer to the $\alpha$-6T (010) oriented surface}
\label{sec:transition}

The transition from the flat adsorption geometry to the crystalline herringbone structure starts already in the second layer. In the following, we compare the structural motifs of the second layer to the features of the $\alpha$-6T bulk structure.\par
The experimental structure of the (010) oriented $\alpha$-6T bulk structure is shown in Figure~\ref{b_vs_c}~(a) and has been obtained from Ref.~\citenum{Horowitz1995}. The $\alpha$-6T (010) terminated crystal exhibits an AB stacking visualized by different colors. Layer A is colored in orange and layer B in blue. The molecules in the second row are depicted by lighter colors. All molecules are tilted by $\phi \approx 33^{\circ}$ with respect to the (010) plane.\cite{Oehzelt2009} The molecules in A and B are tilted in the opposite direction. The vertical translation between two molecules with opposite tilt angles will be further denoted as $\Delta z$ and amounts to $\approx \SI{2.9}{\angstrom}$.\par 
The structural features of the $\alpha$-6T crystal are depicted in more detail in Figures~ \ref{b_vs_c} (b) and (c). Relative shifts between molecules in two consecutive AB layers are observed. The lateral shift, $\Delta_{\textrm{lat}}$, is with $\SI{2.7}{\angstrom}$  very similar to the lateral translation between the first and second layer in the 2S@1S and 2R@1R structures ($\approx \SI{3.0}{\angstrom}$, see Table~\ref{bl_features}).  Furthermore, the molecules in two consecutive layer are shifted along the LMA by $\SI{1.2}{\angstrom}$ with respect to each other as visualized by the blue arrow in Figure \ref{b_vs_c} (c). We explain the significantly larger $\Delta_{\textrm{LMA}}$ shift in the bilayer by the presence of the metallic surface:  a lateral shift of $\approx \SI{3}{\angstrom}$ locates the outermost thiophene unit directly above the gap between two molecular rows in the first layer, enabling some of the molecules to bend the first thiophene unit downwards, see Figures~\ref{bl_shift} (d) and \ref{bl_tilt} (e). Such a downward bent conformation maximizes the interaction with the surface, which is about 10 times stronger than the molecule-molecule interactions, see Ref.~\citenum{Scarbath-Evers2019} and the interaction energies given in Table~S2 in the SI. \par 
Characteristic features of the $\alpha$-6T (010) oriented bulk structure are already present in the 2S and 2R layers, such as the shifts $\Delta_{\textrm{lat}}$ and $\Delta_{\textrm{LMA}}$ and the tilt of the molecules. However, the arrangement of the molecules in the second layer does not yet coincide with that of the molecules in the $\alpha$-6T bulk structure. The structural differences comprise lattice constants, the magnitude of the molecular tilt, $\phi$, and the vertical translation, $\Delta z$, between molecules with opposite tilt. We discuss these differences in the following.\par
For the surface unit cell of the bilayer structures, the measured lattice constants given in Table~\ref{tab1} are a factor of $1.1-1.2$ larger than the lattice constants of the bulk structure. This can be explained by the lattice mismatch between the unit cell of the flat monolayer and the $\alpha$-6T (010) oriented cell. The monolayer acts as a template for the growth of the second layer enforcing a surface unit cell that is larger than the unit cell of the bulk.\par
The tilt angles, $\phi$, of the molecules in the 2S and 2R layer differ in size and orientation from the tilt angle in the (010) oriented crystal. In the bulk structure, layers A and B are tilted in opposite directions. However, the orientation of $\phi$ is the same within the layers,  similar to scenario II depicted in Figure \ref{comic_scenario} (d). As discussed in detail before, the measured and computed  2R$@$1R and 2S$@$1S structures exhibit a staggered arrangement corresponding to structure \RM{3}. Moreover, the tilt angle in the 2S and 2R layer is significantly smaller than the bulk tilt ($ \approx \SI{14}{\degree}$/$\SI{12}{\degree}$ vs. $\SI{33}{\degree}$, see Table~\ref{bl_features}). The alternately tilted arrangement in the 2S and 2R layers resembles a hybrid structure of the bulk layers A and B. This is visualized in Figure~\ref{b_vs_c}~(d) and (e), where the molecules in the 2S and 2R layer are colored in orange and blue to indicate their resemblance to the respective parts in the bulk structure.

\subsection{An unexpected growth mechanism for $\alpha$-6T on Au(100)}
\label{section:growth}
The observation of a staggered structure for the $\alpha$-6T bilayer on Au(100) challenges the common paradigm for the growth of organic crystals on strongly interacting substrates. Strong molecule surface interactions as reported for $\alpha$-6T on Au(100) in Table~S2 (SI) and in Ref.~\citenum{Scarbath-Evers2019}, trigger typically an epitaxial growth of flat layers in a Frank-van der Merwe mode. This growth continues usually for a few layers, but can persist, e.g, for pentacene adsorbed on thermally treated graphene, until a film thickness of 110~nm.\cite{Jo2015} In general, after reaching a critical thickness in the film, a fast transition into the bulk structure occurs. This transition can induce a restructuring of the flat layer, as proposed for para-sexiphenyl (6P) on Ag(100).\cite{Hollerer2018} More commonly, the growth of 3D islands in the relaxed bulk structure is observed on top of the flat layers (Stranski-Krastanov growth). The latter is the case for $\alpha$-6T on Au(111), where the formation of 3D clusters starts after completion of the second layer,\cite{Bronsch2018} but has been also observed for adsorption of 6T and similar molecules on oxide surfaces.\cite{Haber2008,Sun2010}\par
The staggered structure of the $\alpha$-6T bilayer on Au(100) coincides neither with a Frank-van der Merwe or island-based Stranski-Krastanov growth mechanism. A restructuring of the first layer is also not observed. Instead, we propose a mechanism in which $\alpha$-6T gradually approaches its bulk structure over several layers. The aspect of inclination of the molecules in the second layer has also been suggested for pentacene on Cu(111) at higher packing densities; \cite{Smerdon2011} in this case the existence of smaller domains containing tilted molecules has been assumed based on their apparent height. For quaterthiophene ($\alpha$-4T) on Ag(111), the growth of a metastable, disordered phase, that differs from the crystalline phase, on top of well-ordered layers of flat lying molecules has been suggested based on infrared spectroscopy experiments.\cite{Li1995,Umbach1995} \par

\begin{figure}[t]
\centering
\includegraphics[width=0.95\linewidth]{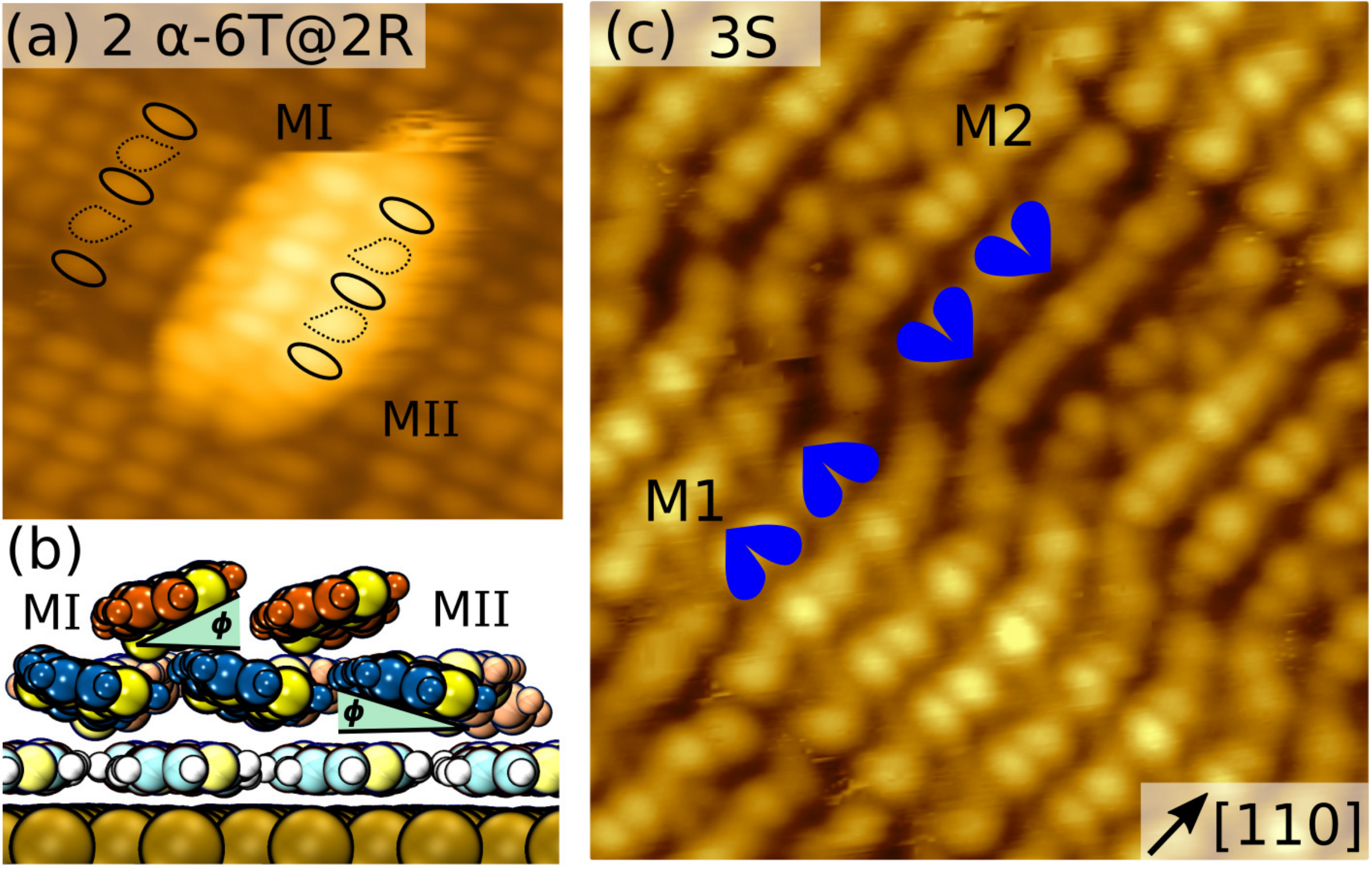}
\caption{(a) Orbital resolved STM image of two $\alpha$-6T molecules adsorbed on a 2R bilayer probing also the deeper valence states (I$_{ \rm{T}} = \SI{5}{\pico\ampere}$,\, U$_{\rm{T}} = \SI{-2}{\V}, \SI{25}{\kelvin}$) (b) Computed structure for two molecules on top of a 2R@1R bilayer. (c) Orbital resolved STM image of two molecular rows in the third layer measured at displaying the HOMO ($\SI{-1.35}{\V},\, \SI{25}{\kelvin},\, \SI{5}{\pico\ampere}$). }
\label{3rd_layer}
\end{figure}

To corroborate our gradual transition model, we investigate the beginning growth of the third layer for two molecules adsorbed on top of a 2R bilayer. The corresponding experimental STM image is shown in Figure~\ref{3rd_layer} (a), where we have introduced the labels M\RM{1} and M\RM{2} for the two adsorbed molecules. The STM image has been measured at a bias voltage of $\SI{-2}{\V}$ and probes also the deeper valence states. The protrusion pattern for the second layer is thus different than described before. Individual molecules appear now as 5 aligned bright spots. The molecules in the beginning third layer, M\RM{1} and M\RM{2}, show a similar brightness pattern compared to the molecules in the second layer underneath. In addition, we observe several darker protrusion on the left of M\RM{1}. The height profile measured vertically along the LMA shows that they are $\approx \SI{1.2}{\angstrom}$ lower in height (see Figure S5 in the SI), which strongly indicates that the molecules in the third layer are also tilted with respect to the surface plane. The darker protrusions can be assigned to thiophene units which are closer to the surface. Both molecules must have the same orientation since the darker protrusions are not visible for M\RM{2}. They are superimposed by the bright protrusions of the adjacent molecule M\RM{1}. These findings are in agreement with the result of our computational structure search shown in Figure~\ref{3rd_layer} (b).  The tilt angle predicted by our computational model for M\RM{1} and M\RM{2} is  with $18 ^\circ$ only slightly larger than for the second layer and therefore still significantly smaller than in the $\alpha$-6T crystal, indicating that the molecules have not yet approached their bulk conformation. \par

 Furthermore, our computational model predicts that the molecules in the beginning third layer are tilted in the opposite direction with respect to the molecules in the row below, leading to an AB-like structure as in the (010) oriented bulk structure. Evidence for the latter is also found from the symmetry and shape of the protrusion pattern in the experimental STM image in Figure~\ref{3rd_layer} (a). Three out of the 5 protrusions per molecule have elliptical shape (solid black line), while the other two appear  droplet-shaped (dashed black line). We find that the protrusion pattern of M\RM{1} and M\RM{2} are 180$^{\circ}$ replica of the ones in the molecular row underneath. Detailed analysis of symmetry arguments presented in the SI (Figure~S6) lead to the conclusion that the 180$^{\circ}$ rotation can be only observed when the molecules are tilted in the opposite directions.  \par

Further evidence for a gradual growth mechanism is obtained from STM studies of the fully formed third layer. Figure~\ref{3rd_layer} (c) shows a high-resolution image of an S-domain on top of an $\alpha$-6T bilayer (3S). Molecules in consecutive molecular rows are again denoted as M1 and M2. For direct comparison to the STM image of the second layer (Figure~\ref{bl_tilt} (a)), the bias voltage has been chosen such that solely the HOMO density is displayed. The protrusion pattern of the third layer strongly resembles the one of the bilayer regarding the number of protrusions, their symmetry and variation in brightness. We can thus conclude that the staggered structure is resumed in the third layer. Additionally, the molecular rows appear alternately brighter and darker, best visible from the large-scale STM image in Figure~S7 (SI), indicating height differences between adjacent rows with opposite tilt. The vertical translation $\Delta z$ between molecules with different orientation is a typical feature of the bulk structure (see Figure \ref{b_vs_c} (a)), which was still absent in the second layer. Its incorporation in the third layer strongly supports our proposed growth mechanism, in which the (010) contact plane of the bulk structure is gradually approached over several layers.


\section{Conclusion}
We employ high-resolution STM measurements in combination with hybrid QM/MM calculations to study
the morphology of $\alpha$-6T on Au(100) beyond monolayer coverage. We observe two chiral domains, 2R and 2S, on top of the flat monolayer domains of 1R and 1S chirality, respectively. Remarkably, adsorption beyond the flat monolayer does not follow the epitaxial flat growth. Instead, the molecules gradually adopt structural motifs of the $\alpha$-6T bulk structure: a lateral shift, a shift along the LMA, and a tilt around the long molecular axis and, for the third layer, a vertical translation between molecules with opposite tilt angles.\par
Despite a strong structural resemblance to the (010) oriented $\alpha$-6T bulk structure, none of the found structures, 2R and 2S, coincides with a well-defined plane of the $\alpha$-6T crystal. Instead, the second layer should be regarded as a structural transition zone where the attempt of the $\alpha$-6T molecules to adsorb with a specific contact plane on the flat monolayer has not yet outbalanced the effect of the metal surface. The structure of the third layer does also not yet coincide with the (010) bulk structure of $\alpha$-6T. These findings suggest an  unexpected growth mechanism in which the transition of the flat monolayer to the staggered bulk structure occurs gradually over several layers in a layer-by-mode.  Due to the strong dependence of the opto-electronic properties on the local morphology, this new growth mechanism is of paramount importance for the application of rod-like molecules in opto-electronic devices.

\section{Experimental and computational details}
All STM measurements have been performed in ultrahigh vacuum conditions at $\SI{80}{\kelvin}$ or at $\SI{25}{\kelvin}$. The Au(100) sample has been prepared by Ar$^{+}$ sputtering and annealing cycles followed the procedure  in Ref. \citenum{Hammer2014a}. For the STM measurements electrochemically etched tungsten tips were used. The $\alpha$-6T molecules were evaporated onto the sample at room temperature by sublimation from a Knudsen cell at a temperature of $\SI{495}{\kelvin}$. A sublimation rate of $0.06$ monolayer per minute allowed the preparation of $\alpha$-6T layers of well-defined thickness. Temperature control during measurement was achieved using a chromel-alumel thermocouple welded by a laser to the Au(100) crystal. The energy of the molecular electronic states (i.e. HOMO, HOMO-1) was obtained from scanning tunneling spectroscopy measurements and we chose the bias voltage of the STM measurement accordingly. \par

We took the hexagonal Au surface as a model system for the reconstructed Au(100) surface. The metallic substrate is modeled by a five-layer slab and laterally by  p$\left(18 \times 8 \right)$ repetition of the unit cell  using the experimental lattice constant of Au (\SI{4.078}{\angstrom}).\cite{Wyckoff1963} Periodic boundary conditions are applied in all three dimensions. To decouple the periodic images in $z$ direction, at least \SI{15}{\angstrom} of vacuum are added. The molecules have been only absorbed on one side of the slab, while the atoms of the lowest three layers are kept fix in their bulk position.  \par
All calculations were carried out with the quantum chemistry package CP2K 4.0.\cite{Hutter2014} In order to tackle the huge system size (up to 1336 atoms per unit cell) we employ a quantum mechanics/molecular mechanics (QM/MM) approach. Recently, such hybrid schemes have been successfully used to study physisorbed and also chemisorbed interfaces.\cite{Golze2013,Golze2015,Rinkevicius2014,Li2014QMMM,Dohn2017,Hofer2015,Rinkevicius2016,Li2016,Saleh2019} We employ an image-charge augmented hybrid model \cite{Golze2013} (IC-QM/MM), which has been specifically developed for adsorbate-metal systems accounting for induction effects by applying the image charge formulation. In the IC-QM/MM approach, the adsorbates are treated quantum mechanically, while the metal atoms and the interactions between the subsystems are treated at the MM level of theory.\par
The QM subsystem ($\alpha$-6T molecules) is calculated with DFT representing the valence electrons by double-$\zeta$ plus polarization basis sets of the MOLOPT type.\cite{Vandevondele2007} In order to describe the interactions between valence and core electrons, norm conserving Goedecker, Teter, and Hutter (GTH) pseudopotentials \cite{GTH1996,GTH1998, Krack2005} were employed. The exchange correlation potential is modeled by the Perdew-Burke-Ernzerhof functional using $\SI{25}{\percent}$ of exact exchange (PBE0).\cite{PBE1996,Adamo1999,Ernzerhof1999} We employed the well-established auxiliary density matrix method (ADMM) \cite{Guidon2010} to reduce the computational cost for the calculation of Hartree-Fock exchange. Inclusion of $\SI{20}-\SI{40}{\percent}$ of exact exchange in the DFT functional is essential for an accurate description of the electronic structure of $\pi$-conjugated organic systems and for an accurate prediction of their molecular arrangement.\cite{Kertesz2005,Choi1997a,Salzner2007a} Highly accurate $G_0W_0$ calculations,\cite{Golze2019} also performed with CP2K\cite{Wilhelm2016,Wilhelm2018} for a thiophene monomer, confirm that the PBE0 functional yields the correct energetic ordering of the HOMO, HOMO-1 and lowest unoccupied orbital (LUMO), see Table S1 (SI). This gives us confidence that the predicted HOMO densities and computed STM images of $\alpha$-6T layers are qualitatively correct. %
Dispersion interactions within the QM subsystem are accounted for by Grimme's D3 correction.\cite{Grimme2010}\par
The MM-based interactions between the Au atoms are described through the embedded atom model (EAM) potential.\cite{Foiles1986} The electrostatic interactions between adsorbates and metal are accounted for by the image charge approach, whereas the dispersion interaction and Pauli repulsion are modeled by a Lennard-Jones potential. The Lennard-Jones parameters (see Table S3 in SI) have been generated from Refs. \citenum{Heinz2008} and \citenum{Moreno2010} using the Waldman-Hagler mixing rules.\cite{Waldman1993} We validated these parameters by comparing the adsorption energies obtained from our IC-QM/MM model to full-DFT calculations, see SI for details.\par
Our structure search started from different initial geometries, including structures with a flat and equally tilted second layer, to ensure that the global minimum has indeed been found.\par

For the STM simulations, the widely used Tersoff-Hamann approximation\cite{Tersoff1985,Tersoff2} is employed to reproduce the isocurrent topography above the second layer at the given bias voltages, as described in detail in Ref.~\citenum{Ding2011}. The tip is modeled by an atomic wave function with $s$ orbital symmetry.
 The computed STM images were analyzed using the scanning probe microscopy software WSXM.\cite{Horcas2007}

The bulk structure of $\alpha$-6T in its low temperature phase has been computed at the DFT level using the PBE functional.\cite{PBE1996} The optimized lattice parameters are given in Table~S4 and the corresponding structural parameters in Table~\ref{bl_features}.\par
All pre-processing and post-processing structural analyses was carried out using the python package MDAnalysis \cite{Michaud-Agrawal2011} and TRAVIS.\cite{Brehm2011}

\section*{Acknowledgement}
This work was supported by the Deutsche Forschungsgemeinschaft (SFB TRR 102, B2, Project-ID 189853844 - TRR 102). L. K. Scarbath-Evers thanks the Center for Information Services and High Performance Computing (ZIH) at TU Dresden for generous allocations of computer time. D. Golze acknowledges support by the Academy of Finland through grant no. 316168.
\section*{Conflicts of interest}
There are no conflicts to declare.

\providecommand{\latin}[1]{#1}
\providecommand*\mcitethebibliography{\thebibliography}
\csname @ifundefined\endcsname{endmcitethebibliography}
  {\let\endmcitethebibliography\endthebibliography}{}

\end{document}